\documentclass[aps,prl,showkeys,superscriptaddress,twocolumn,nofootinbib,floatfix]{revtex4-1}
\usepackage{amsmath}
\usepackage{amssymb}
\usepackage{amsthm}
\usepackage{mathrsfs}
\usepackage{graphicx}
\usepackage{epstopdf}
\usepackage{fancyhdr}
\usepackage{array}
\usepackage[all]{xy}
\usepackage{eufrak}
\usepackage{euscript}
\usepackage{enumerate}
\usepackage{slashed}
\usepackage{hyperref}
\usepackage{caption}
\usepackage{epstopdf} % COMPILE EPS (AND PDF) FIGURES USING PDFLATEX!!!

\newcommand{\be}{\begin{equation}}
\newcommand{\ee}{\end{equation}}
\newcommand{\bea}{\begin{eqnarray}}
\newcommand{\eea}{\end{eqnarray}}

\def \v {\vec}

\begin{document}

\title{Suppression of baryon diffusion and transport in a baryon rich strongly coupled quark-gluon plasma}

\author{Romulo Rougemont}
\email{romulo@if.usp.br}
\affiliation{Instituto de F\'{i}sica, Universidade de S\~{a}o Paulo, C.P. 66318, 05315-970, S\~{a}o Paulo, SP, Brazil}

\author{Jorge Noronha}
\email{noronha@if.usp.br}
\affiliation{Instituto de F\'{i}sica, Universidade de S\~{a}o Paulo, C.P. 66318, 05315-970, S\~{a}o Paulo, SP, Brazil}
\affiliation{Department of Physics, Columbia University, 538 West 120th Street, New York, NY 10027, USA}

\author{Jacquelyn Noronha-Hostler}
\email{jnoronhahostler@phys.columbia.edu}
\affiliation{Department of Physics, Columbia University, 538 West 120th Street, New York, NY 10027, USA}

%\date{\today}

\begin{abstract}
Five dimensional black hole solutions that describe the QCD crossover transition seen in $(2+1)$-flavor lattice QCD calculations at zero and nonzero baryon densities are used to obtain predictions for the baryon susceptibility, baryon conductivity, baryon diffusion constant, and thermal conductivity of the strongly coupled quark-gluon plasma in the range of temperatures $130\,\textrm{MeV}\le T\le 300\,\textrm{MeV}$ and baryon chemical potentials $0\le \mu_B \le 400\,\textrm{MeV}$. Diffusive transport is predicted to be suppressed in this region of the QCD phase diagram, which is consistent with the existence of a critical end point at larger baryon densities. We also calculate the fourth-order baryon susceptibility at zero baryon chemical potential and find  quantitative agreement with recent lattice results. The baryon transport coefficients computed in this paper can be readily implemented in state-of-the-art hydrodynamic codes used to investigate the dense QGP currently produced at RHIC's low energy beam scan.
\end{abstract}

%\pacs{Valid PACS appear here}
% PACS, the Physics and Astronomy Classification Scheme.
% Valid PACS numbers may be entered using the \verb+\pacs{#1} command.

% Use showkeys class option if keyword display desired

\maketitle

%%%%%%%%%%%%%%%%%%%%%%%%%

\noindent \textsl{1. Introduction.} After more than a decade of intense experimental \cite{QGPexp1,QGPexp2,QGPexp3,QGPexp4,Aad:2013xma} and theoretical investigation, it is now widely believed that the strongly interacting quark-gluon plasma (QGP) formed in high energy ultrarelativistic heavy ion collisions at RHIC and LHC behaves as a nearly perfect relativistic fluid \cite{QGP}. This surprising feature of the QGP may be best illustrated by the very small value of its shear viscosity to entropy density ratio, $\eta/s \sim 0.2$, which has been extracted from comparisons between viscous relativistic hydrodynamic modeling and experimental data (for a recent review, see \cite{Heinz:2013th}). Neither ordinary hadronic transport \cite{Demir:2008tr} nor weak coupling calculations \cite{Arnold:2000dr} have been able to explain such a small value for this ratio in QCD, though alternative explanations have been proposed \cite{Liao:2006ry,NoronhaHostler:2008ju,NoronhaHostler:2012ug,Hidaka:2009ma} that rely on the appearance of a different set of degrees of freedom at the crossover transition.

With the advent of the low energy beam energy scan at RHIC, a new region in the QCD phase diagram \cite{Rischke:2003mt,Kogut:2004su} corresponding to a hot and baryon rich QGP is being explored. The ultimate goal of this effort is to find unambiguous experimental evidence for the existence of a critical end point (CEP) \cite{Stephanov:1998dy,Stephanov:1999zu}. While definite experimental proof of a CEP is still lacking (for a different point of view, see \cite{Lacey:2014wqa}), the observed large degree of collectivity displayed by hadrons with low transverse momentum in low energy RHIC collisions \cite{Adamczyk:2012ku} seems to be compatible with the hypothesis that a baryon rich QGP may also behave as a nearly perfect fluid \cite{Denicol:2013nua}. This experimental observation, together with the success of viscous hydrodynamic modeling in high energy collisions where baryon density effects are small, boosted the interest of the theoretical community in applying hydrodynamics to describe the dynamical evolution of the hot and baryon dense strongly interacting QGP formed in these reactions \cite{Monnai:2012jc,Steinheimer:2012bn,Auvinen:2013sba,Steinheimer:2014pfa,Karpenko:2015xea,Floerchinger:2015efa} where effects from a conserved baryon current cannot be neglected. Given the fundamental role played by hydrodynamical transport coefficients in our understanding of the QGP, one expects that the knowledge about the novel transport coefficients associated with baryon transport may play a key role in the comparison of hydrodynamic calculations to the experimental data obtained in low energy collisions. 

In this case, the underlying effective model within which these transport coefficients are to be computed must not only appropriately describe the known equilibrium properties of the QGP at nonzero baryon densities near the QCD crossover transition \cite{Aoki:2006we}, but also necessarily include from the outset its nearly perfect fluidity. Currently, the only consistent framework that is able to deal with the required complexity behind the strong coupling regime of non-Abelian gauge plasmas and their perfect fluid-like features is the holographic gauge/gravity duality \cite{adscft1,adscft2,adscft3}. Indeed, a universal prediction of this framework obtained a decade ago is that nearly perfect fluidity behavior emerges as a universal property of such strongly coupled plasmas and, in fact, $\eta/s=1/4\pi$ for a large class of gauge theories \cite{Kovtun:2004de}. Since then, the gauge/gravity duality has been applied to obtain many important qualitative insights about the strongly-coupled QGP (for a review, see \cite{solana}).

A few years ago it was realized \cite{bulkviscosity} that holography could also be used to shed light on the real time, dynamical properties of the strongly coupled QGP near the crossover phase transition. The general idea put forward in \cite{bulkviscosity,GN} (and also in \cite{Gursoy:2008bu,Gursoy:2009jd} for a pure glue plasma) is that, due to strong coupling effects, many of the properties of the QCD phase transition may be mimicked by black hole solutions of a simple 5-dimensional holographic model that has the metric, $g_{\mu\nu}$, and a scalar field (the dilaton), $\phi$, as the main dynamical degrees of freedom. This type of Einstein-Dilaton bottom-up model not only displays the same thermodynamic properties as the QGP at zero baryon chemical potential $\mu_B=0$ but also has nearly perfect fluidity built-in. Recently, this holographic setup has been used to investigate, in a quantitative manner, the near-crossover behavior of a large set of physical observables \cite{jorge1,jorge2,gubser1,gubser2,jorge3,jorge4,jorge5,electricconductivity,debye,hydrotransport,qnms,EMD+B,Yang:2015bva} (see also the study in the Veneziano limit in \cite{zahed}). While such models may not be directly derived from string theory, a broader view of the validity of the standard holographic dictionary may be invoked, much like as in the case of condensed matter applications \cite{Hartnoll:2009sz}, to motivate the calculation of dynamical properties of the QGP in a regime that had been otherwise unaccessible to other nonperturbative approaches, such as lattice QCD.

This is especially true in the case of the baryon rich QGP formed in low energy heavy ion collisions. In this case, lattice calculations for the temperature ($T$) and baryon chemical potential ($\mu_B$) dependence of transport coefficients may not be available for quite some time and holography is a natural framework to perform such calculations. In fact, this was the motivation behind the study performed in \cite{EMD+mu} where a holographic Einstein-Maxwell-Dilaton (EMD) model was used to understand the energy loss and some thermalization properties of the QGP at nonzero baryon density. This model, originally proposed in \cite{gubser1,gubser2} to study a holographic critical end point, has its parameters fully fixed by the equation of state of the QGP at $\mu_B=0$, as in \cite{GN}. However, a striking finding discussed in detail in \cite{EMD+mu} was that, without the introduction of new parameters, the model (after the necessary revisions to accommodate more recent $\mu_B=0$ lattice data) can also \emph{quantitatively} describe the available lattice data \cite{fodor1} for the equation of state of the QGP near the crossover transition with baryon chemical potentials up to $\mu_B=400$ MeV. Furthermore, our model prediction for the electric conductivity of the QGP at $\mu_B=0$ \cite{electricconductivity} also gives a good description of recent lattice data \cite{Amato:2013naa} for this observable (see Fig. 6 in \cite{Greif:2014oia} for a direct comparison). These findings give a strong indication that this holographic model may be used to make quantitative predictions for the out-of-equilibrium properties of the QGP at moderate values of the baryon chemical potential.

Here we employ the EMD holographic model \cite{gubser1,gubser2,EMD+mu} to make the \emph{first} realistic predictions for the $T$ and $\mu_B$ dependence of the baryon susceptibility, baryon conductivity, baryon diffusion constant, and thermal conductivity of the strongly interacting QGP at the crossover transition. The baryon susceptibility is found to increase with $\mu_B$ while the baryon conductivity is somewhat robust to the presence of a nonzero $\mu_B$, though it displays a nontrivial temperature dependence. On the other hand, the thermal conductivity increases with $\mu_B$. Moreover, we predict that the baryon diffusion constant is reduced with increasing baryon densities, which indicates that diffusive transport in the strongly-coupled QGP is overall suppressed in a baryon rich environment. We also calculate the fourth-order baryon susceptibility at $\mu_B=0$ and find good quantitative agreement with recent lattice data \cite{Bellwied:2015lba} near the crossover transition.

\noindent \textsl{2. Holographic model.} The EMD action reads
\begin{align}
S=\frac{1}{2\kappa^2}\int_{\mathcal{M}_5}d^5x\sqrt{-g}\left[\mathcal{R}-\frac{1}{2}(\partial_\mu\phi)^2-V(\phi) -\frac{f(\phi)}{4}F_{\mu\nu}^2\right],
\label{EMD}
\end{align}
where the gravitational constant $\kappa^2$, the dilaton potential $V(\phi)$, and the Maxwell-Dilaton coupling $f(\phi)$ shall be dynamically fixed by lattice data at $\mu_B=0$.

We look for charged black holes of the form
\begin{align}
ds^2=e^{2A(r)}\left[-h(r)dt^2+d\v{x}^{\,2}\right]+\frac{dr^2}{h(r)},
\label{backgroundansatz}
\end{align}
$\phi=\phi(r)$, and $A=A_\mu dx^\mu=\Phi(r)dt$, where the radial position of the black hole horizon is given by the largest root of the equation $h(r_H)=0$ and the boundary of the asymptotically $AdS_5$ space is at $r\rightarrow\infty$. In \cite{EMD+mu} it was discussed how to obtain numerical solutions for the functions $A(r)$, $h(r)$, $\phi(r)$, and $\Phi(r)$ once $V(\phi)$ and $f(\phi)$ are specified. The equations are solved with a pair of initial conditions corresponding to the values of $\phi$ and the derivative of the gauge field at the horizon. Different choices for this pair of initial conditions translate into different thermodynamical states of the plasma. The far from the horizon, ultraviolet asymptotics are given by \cite{gubser1,gubser2}: $A(r)\approx\alpha(r)$, $h(r)\approx h_0^{\textrm{far}}$, $\phi(r)\approx\phi_A e^{-\nu\alpha(r)}$, and $\Phi(r)\approx\Phi_0^{\textrm{far}}+\Phi_2^{\textrm{far}}e^{-2\alpha(r)}$, with $\alpha(r)=r/\sqrt{h_0^{\textrm{far}}}+A_0^{\textrm{far}}$ and $\nu=4-\Delta$, where $\Delta$ is the scaling dimension of the operator dual to $\phi$. The ultraviolet coefficients $h_0^{\textrm{far}}$, $\Phi_0^{\textrm{far}}$, $\Phi_2^{\textrm{far}}$, and $\phi_A$ enter in the expressions for the gauge field theory observables, such as the temperature, the baryon chemical potential, the entropy density, and the baryon charge density, respectively:
\begin{align}
T&=\frac{\Lambda}{4\pi\phi_A^{1/\nu}\sqrt{h_0^{\textrm{far}}}},\,\,\,
\mu_B=\frac{\Phi_0^{\textrm{far}}\,\Lambda}{\phi_A^{1/\nu}\sqrt{h_0^{\textrm{far}}}},\,\,\,
s=\frac{2\pi\,\Lambda^3}{\kappa^2\phi_A^{3/\nu}},\nonumber\\
\rho&=-\frac{\Phi_2^{\textrm{far}}\,\Lambda^3}{\kappa^2\phi_A^{3/\nu}\sqrt{h_0^{\textrm{far}}}},
\label{thermovariables}
\end{align}
where $\Lambda\approx 831$ MeV is the energy scale conversion factor needed to express in physical units the quantities computed using the black holes (this scale is obtained via matching the dip of the speed of sound in the model with that found on the lattice \cite{hydrotransport,EMD+mu}). The dilaton potential and the gravitational constant are determined by a comparison to lattice data for the $(2+1)$-flavor QCD equation of state at $\mu_B=0$ \cite{fodor1} while the Maxwell-Dilaton gauge coupling is fixed by the lattice data for the baryon susceptibility at $\mu_B=0$ \cite{fodor2}. The results \cite{EMD+mu} are $\kappa^2=12.5$ and
\begin{eqnarray}
V(\phi)&=&-12\cosh(0.606\phi)+0.703\phi^2-0.1\phi^4+0.0034\phi^6,\nonumber\\
f(\phi)&=&\frac{\textrm{sech}(1.2\,\phi-0.69)}{3\,\textrm{sech}(0.69)}+\frac{2e^{-100\,\phi}}{3}.
\label{modelparameters}
\end{eqnarray}
Using \eqref{modelparameters} we numerically generated $\sim 10^5$ black hole solutions modeling a hot and dense plasma for several values of $T$ and $\mu_B$. It was shown in \cite{EMD+mu} that this holographic model is in very good agreement with the lattice data \cite{fodor1,fodor2} for $130\,\textrm{MeV}\le T\le 400\,\textrm{MeV}$ and $0\le \mu_B \le 400\,\textrm{MeV}$. At larger temperatures the holographic model is not applicable for QGP phenomenology (since it is not asymptotically free) and weak coupling calculations provide a better description of lattice data \cite{Haque:2014rua}.

\begin{figure}[h]
\begin{centering}
\includegraphics[scale=0.60]{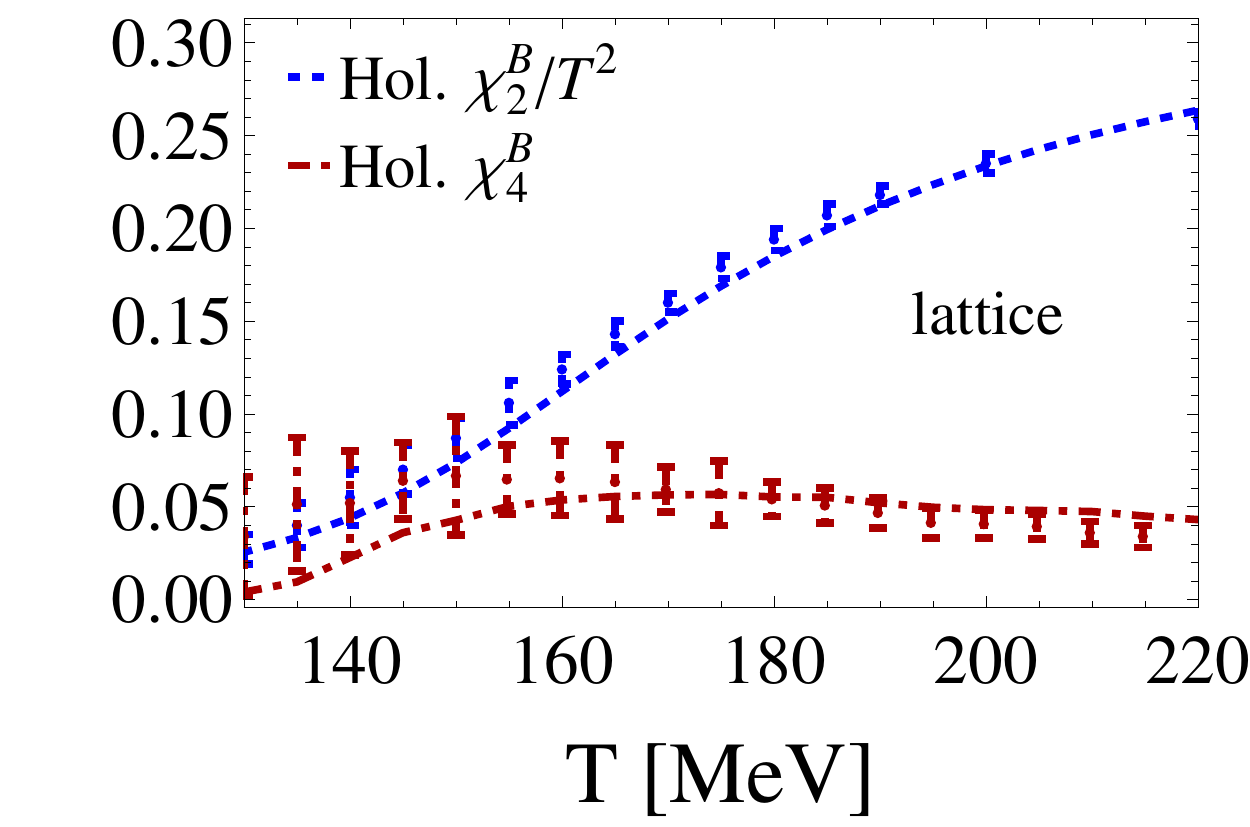}
\par\end{centering}
\caption{(Color online) Second and fourth-order baryon susceptibilities at $\mu_B=0$ in our holographic model compared with lattice data from \cite{Bellwied:2015lba}.}
\label{fig1}
\end{figure}

\noindent \textsl{3. Model predictions.} The $n$-th order baryon susceptibility is given by $\chi_n^B=\partial^np/\partial\mu_B^n=\partial^{n-1}\rho/\partial\mu_B^{n-1}$. In Fig.\ \ref{fig1} we show our prediction for $\chi_4^B$ at $\mu_B=0$ compared with lattice data from \cite{Bellwied:2015lba} (in the same figure we show our fit to the lattice data for $\chi_2^B/T^2$ at $\mu_B=0$, which was used to determine $f(\phi)$ in \eqref{EMD}). One can see that the model can correctly describe the departure of $\chi_4^B$ from $\chi_2^B/T^2$ for $T\geq 150$ MeV, which has been interpreted as a signature of deconfinement \cite{Bazavov:2013dta}; at lower temperatures in the hadronic gas phase the holographic result for $\chi_4^B$ is below the lattice data, which illustrates the domain of validity of the holographic approach.

\begin{figure*}
\begin{center}
\begin{tabular}{c}
\includegraphics[width=0.35\textwidth]{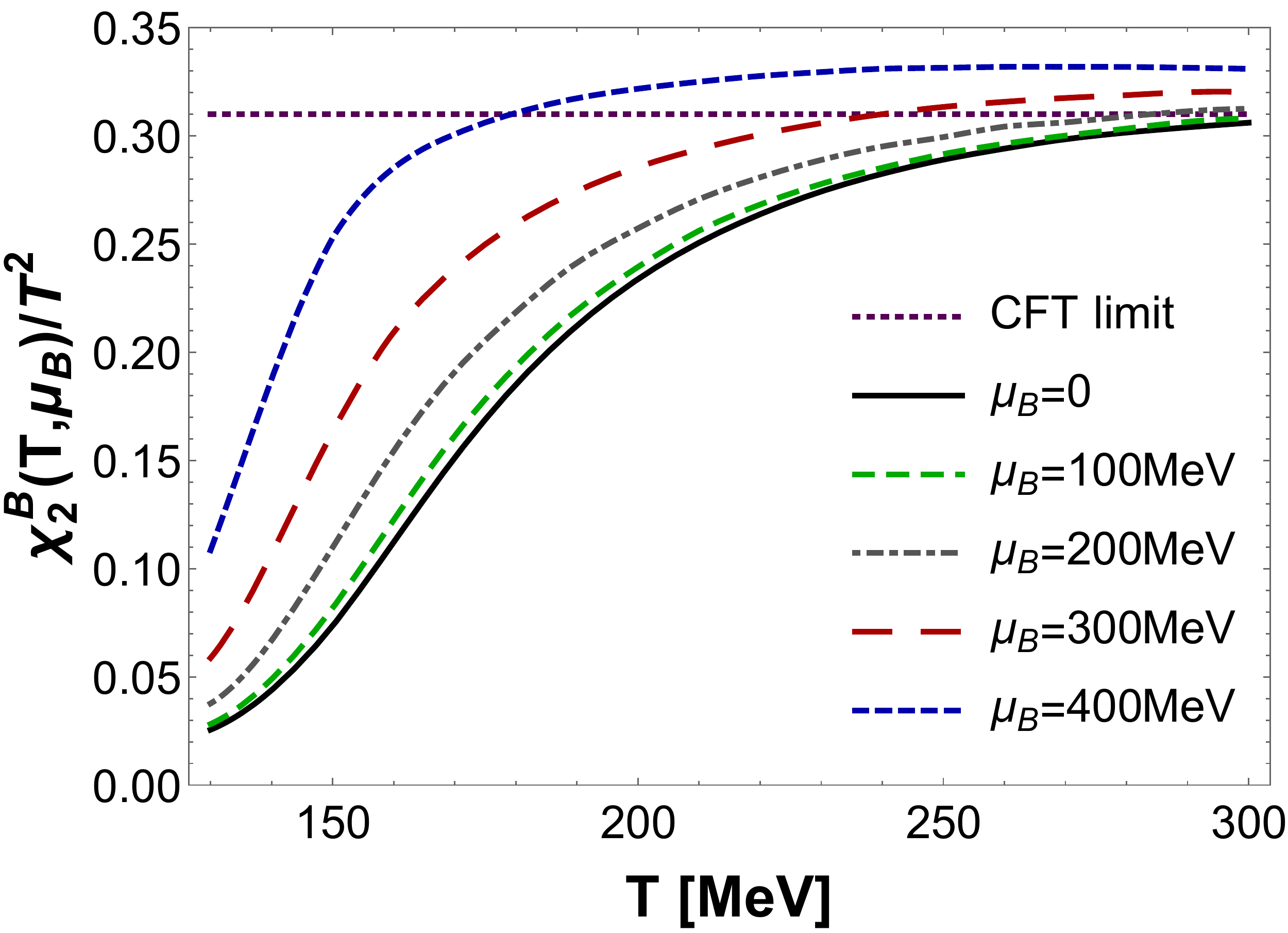} % \\
\end{tabular}
\begin{tabular}{c}
\includegraphics[width=0.35\textwidth]{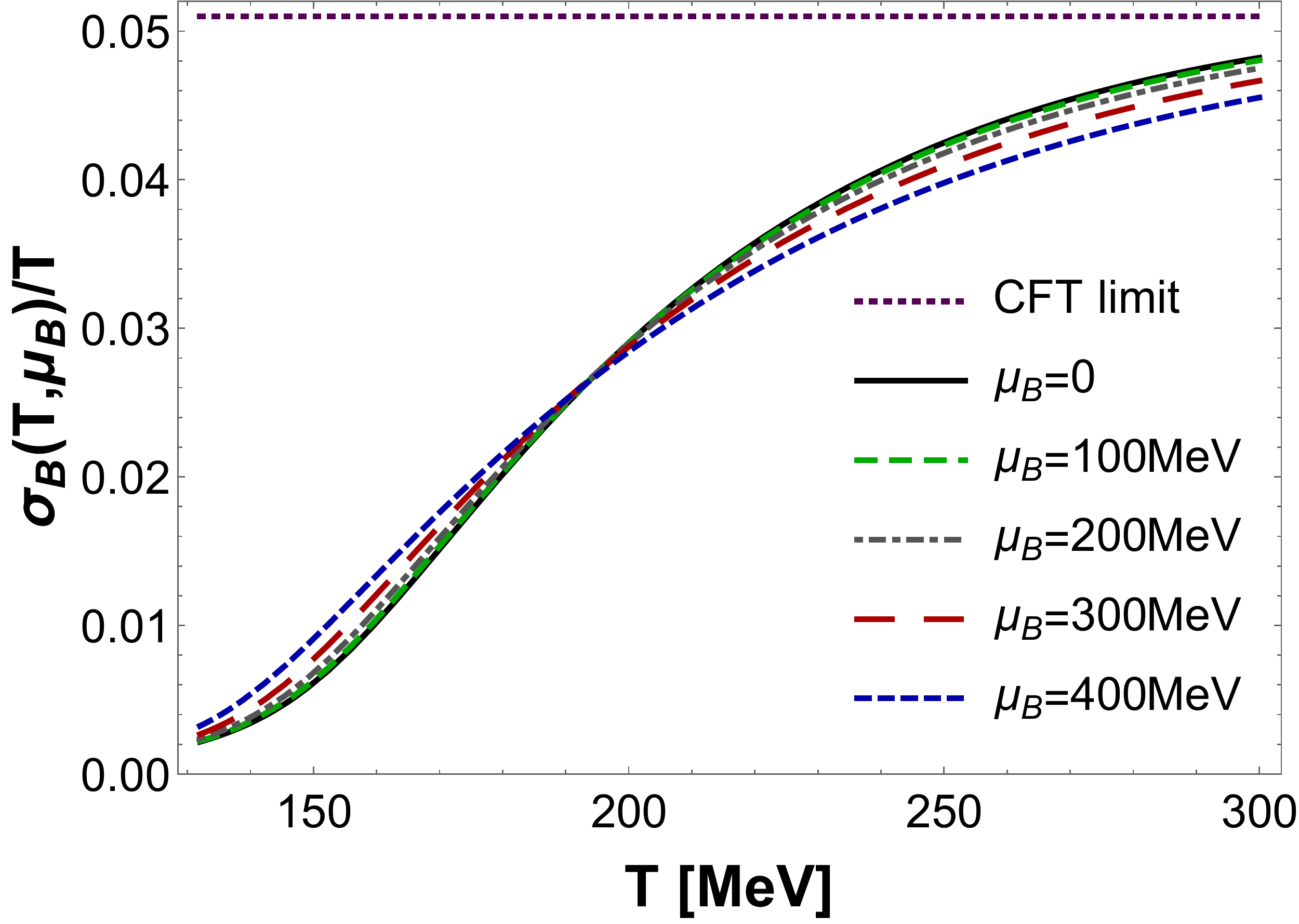} % \\
\end{tabular}
\begin{tabular}{c}
\includegraphics[width=0.35\textwidth]{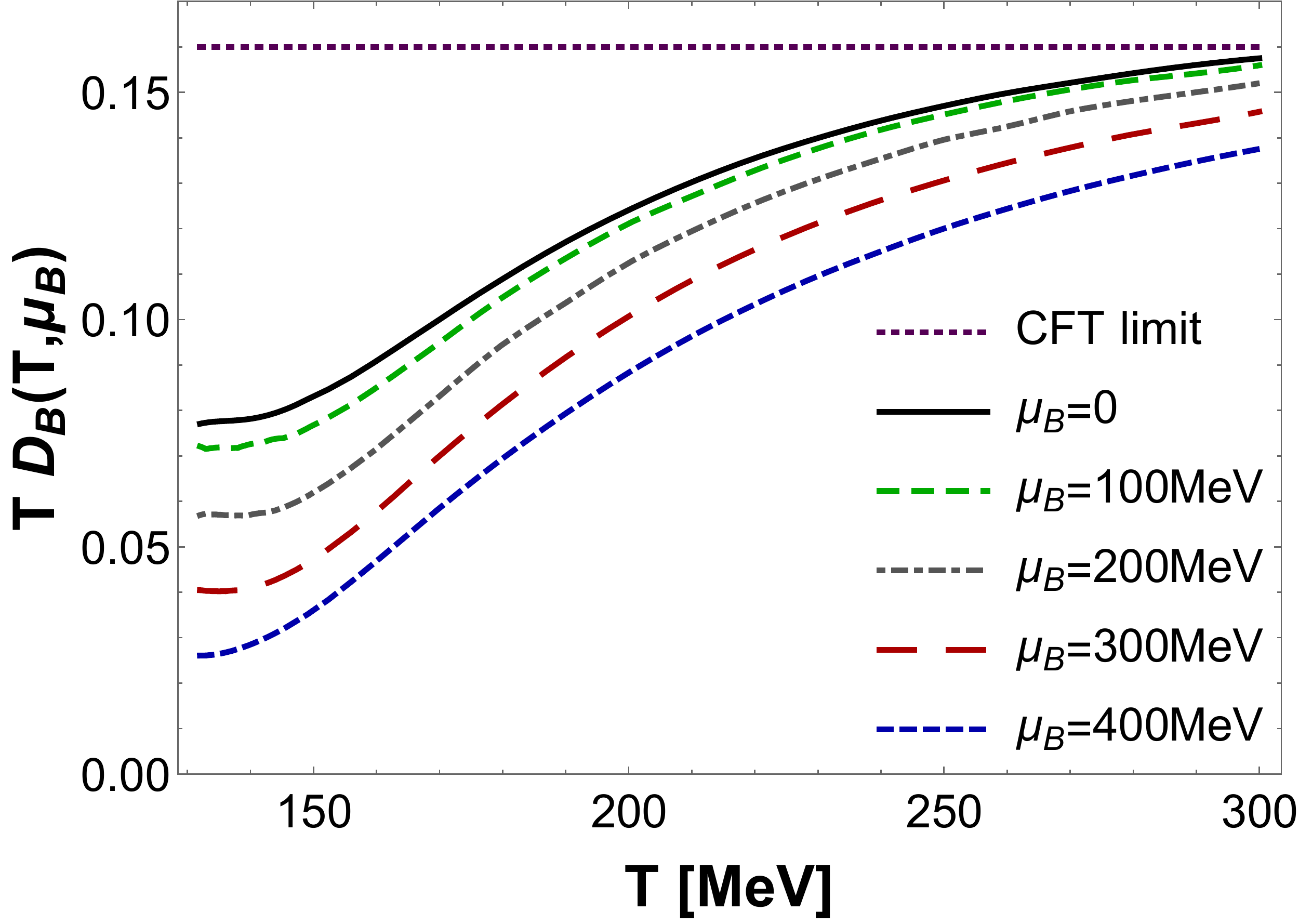} % \\
\end{tabular}
\begin{tabular}{c}
\includegraphics[width=0.35\textwidth]{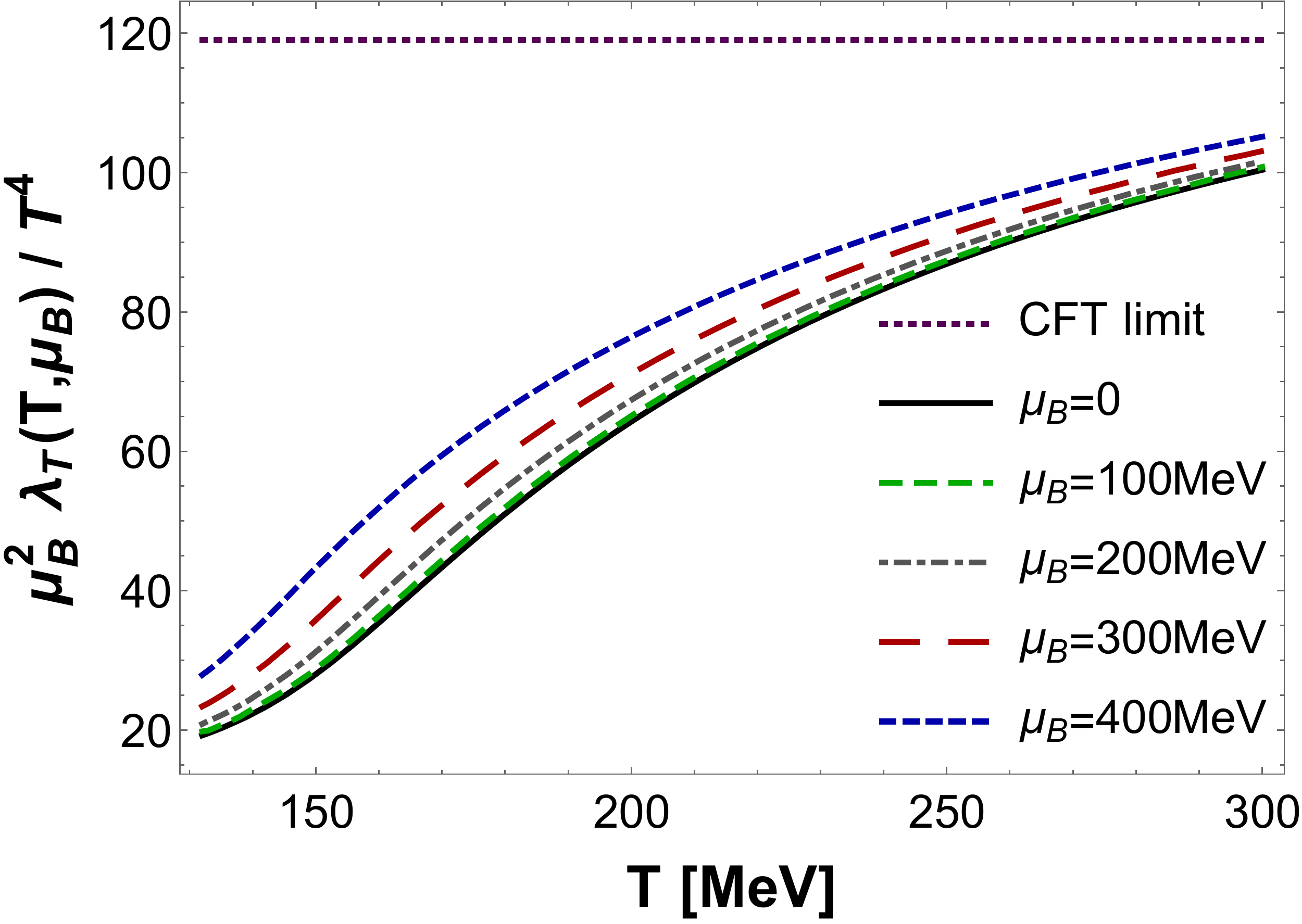} % \\
\end{tabular}
\end{center}
\caption{{\small (Color online) Baryon susceptibility (top left), baryon DC conductivity (top right), baryon diffusion constant (bottom left), and thermal conductivity (bottom right) as functions of the temperature for different values of the baryon chemical potential. The corresponding conformal values attained at large $T$ (and $\mu_B=0$) are also shown.}
\label{fig2}}
\end{figure*}

Our predictions for $\chi_2^B$ at different values of $\mu_B$ are shown in the top left plot in Fig.\ \ref{fig2}. One can see that the baryon susceptibility increases with increasing $\mu_B$. This behavior can be later verified by lattice QCD calculations. We also point out that predictions for the isothermal compressibility, $\nu_T^{-1} = \rho \left(\partial p/\partial \rho\right)_T$, can be directly obtained using our results for $\chi_2^B$ and $\rho$ via $\nu_T = \chi_2^B/\rho^2$ \cite{Kapusta:2012zb}.

We employ linear response theory to compute the baryon conductivity in a holographic setting \cite{PSS}. One considers linearized perturbations around the background \eqref{backgroundansatz} and calculates retarded 2-point functions of gauge and diffeomorphism invariant combinations of these perturbations, which are then associated with transport coefficients via Kubo formulas. As discussed in \cite{gubser2}, due to spatial isotropy each spatial component of the Maxwell perturbation, $a_i,\,i=x,y,z$, satisfies the same decoupled equation of motion and we focus on the $\hat{x}$-direction (the same result will also hold for the $\hat{y}$ and $\hat{z}$-directions), $a\equiv a_x$. Taking a plane wave Ansatz for the perturbation with frequency $\omega$ and zero spatial momentum, which is sufficient for calculating the DC conductivity (i.e, the $\omega\to 0$ limit of the spatially uniform conductivity tensor), one obtains the equation of motion \cite{gubser2}
\begin{align}
a''(r)&+\left(2A'(r)+\frac{h'(r)}{h(r)}+\frac{f'(\phi)}{f(\phi)}\phi'(r)\right)a'(r)\nonumber\\
&+\frac{e^{-2A(r)}}{h(r)}\left( \frac{\omega^2}{h(r)}-f(\phi)\Phi'(r)^2 \right)a(r)=0.
\label{perturbationeom}
\end{align}
The Kubo formula for the DC conductivity in physical units is (excluding the delta function present in translationally-invariant systems at finite density \cite{hartnoll})
\begin{align}
\sigma_B= -\frac{\Lambda}{2\kappa^2\phi_A^{1/\nu}}\lim_{\omega\rightarrow 0}\frac{h(r)f(\phi)e^{2A(r)}\textrm{Im}\left[a^*(r,\omega) a'(r,\omega) \right]}{\omega},
\label{condformula}
\end{align}
where the expression $hf(\phi)e^{2A}\textrm{Im}\left[a^*a'\right]$ corresponds to a radially-conserved flux such that \eqref{condformula} may be evaluated at any value of the radial coordinate, $r$. In \eqref{condformula}, following the prescription proposed in \cite{retarded} to calculate retarded propagators in holography, we consider on-shell configurations of the vector perturbation corresponding to in-falling modes at the black hole horizon which are normalized to unity at the boundary. Then, one may solve \eqref{perturbationeom} for a large grid of initial conditions, plug the solutions with the above specified boundary conditions in the Kubo formula \eqref{condformula}, and obtain the baryon DC conductivity as a function of temperature and baryon chemical potential. 

The corresponding predictions for this transport coefficient are displayed in the top right plot in Fig.\ \ref{fig2}. We find that the dense QGP at the phase transition is not a good baryon conductor. For instance, in the region of the $(T,\mu_B)-$plane considered here, the ratio $T^2 \sigma_B/\eta < 0.05$ in our calculations, which is roughly an order of magnitude lower than recent kinetic theory results \cite{Jaiswal:2015mxa}. One can see that $\sigma_B/T$ varies significantly with $T$ in this regime and that a simple conformal approximation (the straight line in this plot) would completely miss the important effects of the crossover transition. One can also see that above $T\sim 190$ MeV the baryon conductivity is slightly reduced as one increases the chemical potential with the opposite behavior being observed below that temperature (the same qualitative behavior was found in \cite{gubser2}). 

More importantly, the overall dependence of $\sigma_B/T$ with $\mu_B$ is relatively small even though this class of models displays a CEP at larger densities. This is in agreement with the fact that holographic models generally correspond to the type B dynamic universality class \cite{Hohenberg:1977ym} as pointed out in \cite{Natsuume:2010bs} and, thus, this transport coefficient should remain finite as one approaches a CEP. Even though QCD has been argued to belong to the type H dynamic universality class \cite{Son:2004iv}, for the values of $\mu_B$ considered in this work (that are not yet in the critical region) there may not be much difference between H and B classes \cite{Natsuume:2010bs} when it comes to the fluid properties of the system. In fact, the expected dynamical critical exponent for the shear viscosity in QCD $x_\eta \sim 0.054$ \cite{Son:2004iv} is small, which will not spoil the perfect fluidity of the QGP in low energy collisions. Thus, holographic models in which $\eta/s$ does not show singular behavior \cite{Natsuume:2010bs} may be applicable to study the perfect fluid transport properties of the QGP at not so large baryon densities (outside the critical region). In fact, we note that $\chi_2^B$ diverges at the CEP while the baryon diffusion constant (to be discussed below) vanishes for both H and B classes.

As shown in \cite{Iqbal:2008by}, the baryon diffusion may be calculated using the Nernst-Einstein's relation, $D_B = \sigma_B/\chi_2^B$. Our results for this transport coefficient are shown in the bottom left plot in Fig.\ \ref{fig2}. We note that baryon diffusion already at $\mu_B=0$ is predicted to vary significantly across the crossover transition approaching its conformal limit \cite{PSS} only at large temperatures. Furthermore, this plot shows that there should be a significant suppression of baryon diffusion in a baryon rich plasma, even before one reaches a (putative) CEP. This precocious suppression is a consequence of the robustness of $\sigma_B$ to the presence of a nonzero $\mu_B$ and the enhancement of $\chi_2^B$. 

%It cannot be obtained in standard kinetic theory calculations since a proper account of the QCD crossover transition is needed in the determination of baryon transport phenomena.

The bottom right panel in Fig.\ \ref{fig2} shows our results for the thermal conductivity at the crossover transition for several values of $\mu_B$. Following \cite{Kapusta:2012zb}, this may be calculated as $\lambda_T = \left(\sigma_B/T\right)\left[(\epsilon+p)/\rho\right]^2$. One can see that this transport coefficient increases with $\mu_B$ and that its conformal limit \cite{Son:2006em} is only reached at temperatures $T\gg 300$ MeV.

Finally, we point out that in hydrodynamic applications a convenient expression for the dissipative current is $\mathbf{J} = -\kappa_B\nabla  (\mu_B/T)$ with $\kappa_B = T \sigma_B$ \cite{Kapusta:2014dja} and, thus, the results for the baryon conductivity presented here can be readily used to investigate baryon transport effects on the QGP in realistic hydrodynamic calculations (tabulated data for $\sigma_B(T,\mu_B)$ follow attached as an ancillary file).

\noindent \textsl{4. Conclusions.} In this work, the baryon transport properties of a baryon rich QGP near the crossover phase transition were computed holographically using 5-dimensional black holes which quantitatively mimic the thermodynamics of $(2+1)$-flavor lattice QCD data at zero and nonzero baryon chemical potential \cite{EMD+mu}. We made predictions for the baryon susceptibility, the baryon DC conductivity, the baryon diffusion constant, and the thermal conductivity of the strongly coupled quark-gluon plasma in the range of temperatures $130\,\textrm{MeV}\le T\le 300\,\textrm{MeV}$ and baryon chemical potentials $0\le \mu_B \le 400\,\textrm{MeV}$. Baryon diffusive transport is predicted to be precociously suppressed in this region of the QCD phase diagram even though a critical end point could only appear in this model at much larger baryon densities. We also calculated the fourth-order baryon susceptibility at zero chemical potential and found quantitative agreement with recent lattice results around the crossover region. The transport coefficients computed in this paper can be readily implemented in state-of-the-art hydrodynamic models used to investigate the dense QGP currently produced at RHIC's low energy beam scan or at the CBM experiment at the future FAIR facility at GSI.

\noindent \textsl{Acknowledgments.} We thank G.~S.~Denicol and M.~Martinez for discussions about baryon transport. This work was supported by Funda\c c\~ao de Amparo \`a Pesquisa do Estado de S\~ao Paulo (FAPESP) and Conselho Nacional de Desenvolvimento Cient\'ifico e Tecnol\'ogico (CNPq). J.~N.~H. acknowledges support from the US-DOE Nuclear Science Grant No. DE-FG02-93ER40764. J.~N. thanks the Physics Department at Columbia University for its hospitality.

%\appendix
%\section{Appendix A title}
%\label{apa}

\end{document}